\documentclass[prl,aps,twocolumn,superscriptaddress,preprintnumbers,showpacs,floatfix,amsmath,amssymb]{revtex4}

\usepackage{graphicx}
\usepackage{dcolumn}
\usepackage{bm}

\usepackage{amsmath}
\usepackage{graphicx}
\usepackage{amssymb}
\usepackage{mathrsfs}
\usepackage{bbm}
\usepackage{epsfig}

\newcommand{\be}{\begin{eqnarray}}
\newcommand{\ee}{\end{eqnarray}}

\begin{document}

%
\title{ Soliton driven relaxation dynamics and universality in  protein collapse  
}

\author{Andrey Krokhotin}
\email{Andrei.Krokhotine@cern.ch}
\affiliation{Department of Physics and Astronomy, Uppsala University,
P.O. Box 803, S-75108, Uppsala, Sweden}
\author{Martin Lundgren}
\email{Martin.Lundgren@physics.uu.se}
\affiliation{Department of Physics and Astronomy, Uppsala University,
P.O. Box 803, S-75108, Uppsala, Sweden}
\author{Antti J. Niemi}
\email{Antti.Niemi@physics.uu.se}
\affiliation{
Laboratoire de Mathematiques et Physique Theorique
CNRS UMR 6083, F\'ed\'eration Denis Poisson, Universit\'e de Tours,
Parc de Grandmont, F37200, Tours, France}
\affiliation{Department of Physics and Astronomy, Uppsala University,
P.O. Box 803, S-75108, Uppsala, Sweden}

\begin{abstract}
\noindent
Protein collapse can be viewed as a  dynamical phase transition, during which
new scales and collective variables  become excited while the old ones recede and fade away.  
This causes formidable computational  bottle-necks in approaches
that are based on atomic scale scrutiny.  Here we 
consider  an effective dynamical Landau theory to model the folding process at biologically 
relevant time and distance scales.  We reach both a substantial decrease 
in the execution time and improvement in the accuracy of the final configuration, 
in comparison to more conventional approaches.
As an example we inspect  the collapse of HP35 
chicken villin headpiece subdomain, where there are detailed  molecular dynamics 
simulations to compare with.  We start from a structureless, unbend and 
untwisted initial configuration. In less than one second of wall-clock time on a single processor 
personal computer we  consistently reach  the native state with  0.5 \.Angstr\"om root 
mean square distance (RMSD)  precision.  We confirm that our folding pathways are indeed
akin those obtained in recent atomic level molecular dynamics simulations. We conclude that our approach 
appears to have the potential for a computationally economical method to
accurately understand  theoretical  aspects of protein collapse.
\end{abstract}



\maketitle

\section{Introduction}

Structural classification shows  that folded proteins  in the Protein Data Bank (PDB)  \cite{pdb}
are built in a modular  fashion  from a relatively small number of components \cite{comp}. In 
SCOP \cite{scop} there are presently 1393 unique folds
while CATH  \cite{cath} has 1282 different 
topologies. Both figures have remained unchanged since year 2008.  Furthermore,
according to  \cite{xubiao} over 90$\%$ of all high resolution PDB proteins can be modeled 
using no more than 200 explicit soliton motifs  as the modular  blocks. 
This convergence in  protein architecture proposes that in the vicinity of the native state the 
atomic level differences between amino acids become less important in determining the fold.  
Instead the protein shape is dominated collectively, by interactions between a relatively small number of
modular components that are made of several amino acids. 

Classical molecular dynamics (MD)  \cite{md}-\cite{gro} remains the only viable  approach to
describe truly atomic level protein dynamics.  With the best available precision MD models
very  short time and distance  scale oscillations of  individual atoms, including both
their small amplitude  thermal fluctuations and the detailed interactions between  all the
different atoms. But the process of protein folding
engages various different temporal and spatial scales. In particular there are several 
high energy barriers that can be overcome only by relatively slow and 
collective long range oscillations. These hurdles in scales and structures
are the major technical bottle-necks in full atomic level descriptions of protein folding.  
As a consequence a detailed MD simulation of an entire folding process remains a 
formidable  task.   Even in the case of relatively simple and
short proteins such as the  35-residue subdomain of the villin headpiece (HP35) where detailed information on the
folding dynamics is now becoming available,
a detailed simulation can take several months and even  years  to complete \cite{pande}-\cite{fred2}.

In order to enable practical modeling of the folding, several different effective approaches
have been introduced \cite{fred2}. Examples range from the coarse-grained Go model and its  variants
\cite{go1}  to carefully crafted energy functions such as UNRES \cite{sche} that explicitely aim to
average over those degrees of freedom that are considered to be non-essential for attaining
thermodynamically stable structures. 

In addition to technical issues,  there are also important conceptual matters that need to be addressed in
selecting the pertinent coarse grained physical degrees of freedom. In particular, a folding pathway from  a structureless straight protein backbone 
into a biologically active collapsed conformation involves a phase transition. During a phase transition
a physical system undergoes a drastic metamorphosis, old factors  loose their relevance while new 
actors enter the stage.  In the case of  protein folding, the highly localized and short time scale atomic 
oscillations become replaced by much slower collective motions of extended modular structures.
We propose to overcome this dual problem of phases and scales 
in terms of an effective dynamical Landau-type  theory. In analogy of {\it e.g.} the effective Landau-Ginzburg theory 
that models a superconductor in terms of collective Cooper pairs and vortex lines in lieu of the individual 
electrons and photons of the microscopic BCS theory, we aim to describe the dynamics of protein collapse
as a non-conservative  Markovian relaxation process of the relevant modular components. 
In our  approach, a protein consistently folds to its PDB structure with a  
subatomic precision that matches and even exceeds the experimentally 
determined B-factor Debye-Waller fluctuation distances.  Since our description only 
engages those time and distance scales that characterize biologically relevant 
motions, we can reach a  sub-second  execution time of the entire folding process
even with an ordinary desktop computer.

\section{model}

The effective Landau energy is \cite{oma1}, \cite{ulf}
\[
E = \sum\limits_{i=1}^{N}  \biggl\{ 2 a  \kappa_{i+1} \kappa_i   + 2 \kappa_i^2 + b \kappa_i^2 \tau_i^2 + c
( \kappa_i^2 - m^2)
\biggr\} 
\]
\begin{equation}
\ + \ \sum\limits_{i=1}^N 
\biggl\{  d \tau_i + q \kappa_i^2 \tau_i + e  \tau_i^2
\biggr\} 
\label{E1}
\end{equation}
Here $\kappa_i$ and $\tau_i$ are the Frenet bond and torsion angles of the C$_\alpha$ backbone, the summation
extends over all   backbone angles ($\kappa_{N+1}=0$), see \cite{frenet} for details. Once these angles are known the protein backbone can be constructed by 
solving the discrete Frenet equation \cite{frenet}. 
The first sum  defines  the Hamiltonian of the discrete nonlinear 
Schr\"odinger equation \cite{fadde}. In the second sum the first two terms are the 
conserved helicity and momentum, respectively.  
The last term is the Proca mass.  The functional form  (\ref{E1}) is firmly anchored  in the 
elegant mathematical structure of integrable models \cite{fadde},  it describes the protein 
backbone in terms of universal physical arguments \cite{oma1}.  
The various parameters have constant values over each of the soliton motifs  {\it i.e.}  they  
are characteristic only to an entire supersecondary structure such as a helix-loop-helix \cite{cherno}.  

In \cite{nora} it has been shown that (\ref{E1}) supports  solitons as classical solutions.
In \cite{xubiao} it has been shown that over 90$\%$ of all PDB proteins with resolution better 
than 1.5 \.A can be described in a modular fashion in terms of 200 explicitely constructed
soliton profiles. The soliton emerges as follows: We first eliminate the variable $\tau_i$
in favor of $\kappa_i$,
\begin{equation}
\tau_i [\kappa_i] =  - \frac{1}{2}  \frac{d + q \, \kappa_i^2}{ e + b\, \kappa_i^2} 
\label{Etau}
\end{equation}
If the value of $\tau_i$ falls outside of the fundamental domain $[-\pi ,\pi]$  we redefine it modulo $2\pi$.  
Using the relation (\ref{Etau}) and selecting $a = -1.0$ in (\ref{E1}) we then get
\begin{equation}
\kappa_{i+1} - 2 \kappa_i + \kappa_{i-1} \ = \ U' [\kappa_i] \kappa_i  \ \equiv\ \frac{dU[\kappa]}{d\kappa_i^2} 
\ \kappa_i \ \ \ \ (i=1,...,N)
\label{Ekappa}
\end{equation}
where we define $\kappa_{0} = \kappa_{N+1} = 0$.  This is a generalization of the DNLS equation with
\begin{equation}
U[\kappa] = -   \left( \frac{bd - eq}{2b}\right)^2 \! \! \cdot  \frac{1}{e + b\,\kappa^2}    - 
\left( \frac{ q^2 + 8bcm^2}{4b}\right) 
\cdot  \kappa^2 + 
c \cdot \kappa^4
\label{U}
\end{equation} 
With different parameter values its (dark) soliton describes various protein conformations \cite{xubiao}. 
But as it stands, the soliton models only static PDB configurations. We now wish to
extend this approach into a description of the actual dynamical process of protein folding. Starting 
from an unbiased  initial configuration
we aim to reach a soliton arrangement that models  the desired static PDB protein 
with sub-\.Angstr\"om accuracy. 

We shall initiate our simulations  with a structureless configuration,  
an unbent and untwisted  backbone. This configuration resides in the phase where the radius of gyration
scales with Hausdorff  dimension  $d_H = 1$.   In (\ref{E1}) the parameter $c$ characterizes the 
average strength of hydrogen bonds along the backbone. The ensuing  contribution to energy is largely 
responsible to the formation of regular secondary structures and the subsequent transition into the 
collapsed $d_H \approx 1/3$ phase.  To conform with  our initial configuration we start the simulation 
by setting all $c=0$ initially.  During the  early stage of the simulation we then introduce the hydrogen 
bond interactions by swiftly increasing these parameters  to their final values, in a uniform manner.  At that moment  
we observe the initial formation of regular secondary structures such as $\alpha$-helices and $\beta$-strands.
This is quickly followed by soliton (loop) formation, either by local soliton pair production or by soliton transport 
throught global deformations. This
effectuates a rapid collapse into a molten globule  {\it i.e.} a configuration in the $d_H\approx 1/3$ phase
that then proceeds more slowly towards  the native state. 

As in any phase transition simulation, we need to avoid supercooling that may critically slow 
down the simulation.  For this we utilize  the parameter $a$ in  (\ref{E1}). Depending on sign, we interpret it 
as either a  ferromagnetic or  an antiferromagnetic  coupling along a continuous spin Ising chain of
the  $\kappa_i$.  The antiferromagnetic order models 
the folding nuclei that initiate the phase transition, these are essentially the soliton centers. 
We choose the uniform ferromagnetic
coupling  $a=-1.0$  for all except those bonds where we  foresee the eventual 
location of the center of a soliton.   At the putative soliton locations we introduce an  
initial repulsive antiferromagnetic coupling with $a=+1.0$.  
At the first stages of the simulation, in parallel  with  the introduction of  the hydrogen bond interactions,
we remove the folding nuclei by decreasing the strength of the antiferromagnetic couplings 
so that we arrive at the uniform value $a\equiv -1.0$ along the entire backbone. During the entire 
folding process all other parameters remain intact.
With this initial preparation of the hydrogen bond interactions and with the transient
introduction of folding nuclei in our otherwise homogeneous backbone, we 
avoid supercooling into misfolded states with their  
misplaced or extraneous solitons and soliton-soliton pairs.  Even if such 
states are metastable with an energy (\ref{E1}) that exceeds the energy of the 
native state and eventually decay,  they can have  a very long lifetime 
and  substantially slow down the simulations. 
We remind that in actual proteins 
the hydrogen bonds are similarly produced  during the folding process. There are also
natural inhomogeneities in the amino acid structures  that act as folding nuclei. 
Proline is a good example. 

We propose that during biologically relevant  temporal scales the conformational changes
that drive the collapse can be described in terms of  an appropriate dynamical universality class. 
For this we thermally average over all those very short time scale oscillations and tiny fluctuations of  individual atoms
that are irrelevant to the way how the folding progresses over a biologically relevant time period.
The simplest and by far the most natural choice is to utilize a  Markovian 
Monte Carlo time evolution with the standard, universal  heat bath  probability distribution \cite{glauber}, \cite{lebo}
\begin{equation}
\mathcal P = \frac{x}{1+x} \ \  \ \ {\rm with}  \   \ \ \ x =     \exp\{ - \frac{ \Delta E}{kT} \}  
\label{P}
\end{equation}
Here $\Delta E$ is the energy difference between consecutive MC time steps that we compute from (\ref{E1}).
We choose the numerical value of $kT$ so that we are in the collapsed $d_H \approx 1/3$ phase. We have made 
runs at several different values of $kT$ to confirm that there are no qualitative changes in our results, as long as
$d_H \approx 1/3$. 
Since  (\ref{E1})  is an  approximation of the thermodynamical free energy,  
the parameters in (\ref{E1}) are {\it a priori} temperature dependent  and
at the moment we (still) lack a direct relation between
$kT$ and the physical temperature. 

During the time evolution we 
suffocate any potential rearrangement of covalent bonds along the backbone.
For this  we introduce a self-avoiding 
condition \cite{ulf} that  ensures that  during the folding process the distance between any two backbone 
sites remains  at least as large as the length of a backbone covalent bond.

We emphasize that (\ref{P}) does not  describe the atomic level details of the folding process.
Such details  are  highly sensitive to the initial configuration including solvent and other 
environmental factors,  to the extent that  detailed knowledge of 
a particular atomic trajectory during the collapse can hardly have any real 
meaning. Instead the evolution determined by  (\ref{P}) describes the universal statistical aspects  of trajectories  over biologically relevant scales, 
how the protein backbone proceeds during the dynamical phase transition
from a general class of initial configurations  towards its native state.

\section{example}

As an example we consider the folding dynamics of the 35-residue subdomain of the villin headpiece (HP35).  
The villin  is a small ultrafast folding protein that is 
subject to intense studies by experiments, theory and simulations. The PDB code we use is 1YRF, 
it describes the crystallographic structure at 95K with 1.07 \.A resolution \cite{1yrf}.
In Table I we list the relevant parameter values in (\ref{E1}) together with the corresponding backbone sites.
{\small
\begin{table}[htb]
       \centering
\begin{tabular}{|c|c|c|}
\hline  parameter & soliton-1 & soliton-2    \tabularnewline
\hline \hline
$a$ & 1.0 & 1.0 \tabularnewline 
\hline
$c_1$ & 0.459712  &   0.995867 \tabularnewline
\hline
$c_2$   & 4.5533320 &  9.408796  \tabularnewline
\hline
$m_1$ & 1.504535 & 1.550322  \tabularnewline
\hline
$m_2$  & 1.512836  &  1.535081   \tabularnewline
\hline
$b_\tau$ & - 9.575214e-9 & -1.215692e-08   \tabularnewline
\hline
$d_\tau$ & -6.76965e-11 &  -7.840467e-08  \tabularnewline
\hline
$e_\tau$ & 2.4378718e-8 &  2.136684e-08  \tabularnewline
\hline
$q_\tau$ & 6.769649e-10 &  4.973244e-12   \tabularnewline
\hline
RMSD (\.A) & 0.38 & 0.32 \tabularnewline
\hline
\end{tabular}
\caption{\small
\it Parameter values for the two-soliton solution of (\ref{Etau}), (\ref{Ekappa}) 
that describes the backbone of 1YRF with a combined 0.38\.A accuracy.  
The soliton-1 is located between sites 45-57 (PDB indexing) and the soliton-2 is
located between PDB sites 58-73.  Note that the definition of bond angle $\kappa_{i,i+1}$
takes three and the definition of torsion angle $\tau_{i,i+1}$  takes four sites.
 }
       \label{tab:para}
\end{table}
}
We have determined
the parameters by solving (\ref{Ekappa}), (\ref{Etau}) to describe the folded structure of  1YRF with 
an overall 0.38 \.A RMSD accuracy.

We start the folding simulation with an initial configuration that is straight line, with all $\kappa_{i,i+1}$ and all 
$\tau_{i,i+1}$ equal to zero. This is a configuration with Hausdorff 
dimension $d_H = 1$, and it minimizes (\ref{E1}) for $c=0$ {\it i.e.}
when there are no hydrogen bond interactions. We turn on the hydrogen bonds by increasing   $c$ to the values in
Table I during the first 700.000 steps. We also  introduce the
folding nuclei  by selecting  the initial values $a=+1.0$ for  two bonds that are located between sites 
53-54 and 61-62 corresponding to the centers of the two solitons in 1YRF.
We convert these antiferromagnetic couplings into ferromagnetic $a=-1.0$
in tandem with turning on the hydrogen bonds $c$.  
After this initial preparation we have a random coil configuration.  There is a rapid formation of regular secondary structures 
and  a collapse  into a molten globule,  followed by a relatively slow progress towards the final PDB configuration.  

Since we can reach the final configuration  in less than one second of total execution time 
using a single core in a MacPro desktop computer, we  are able to collect a large
amount of statistical data to investigate the universal aspects of folding pathways.
We find that the folding proceeds in a very universal manner,  the variations between different runs are very small
and the collapse proceeds systematically through steps that are in 
line with the MD simulation in \cite{duan}.  The collapse starts with the formation of the last helix III. The second
loop and the middle helix II then appear, followed by the formation of  helix I and the first loop. 
Towards the end of the collapse the helix I starts to stabilize at a slightly faster  rate than the middle helix II.  During the 
final stages  the folding process consists mainly from an adjustment  of the middle helix II with the
two adjacent loops. This helix formation is very similar to the observations made during MD simulations
as reported in \cite{wallin}. This can
be seen by comparing our Figure 1 with the corresponding Figure in \cite{wallin}.
%
%
%
%
%
\begin{figure}[tbh]
        \centering
                \includegraphics[width=0.45\textwidth]{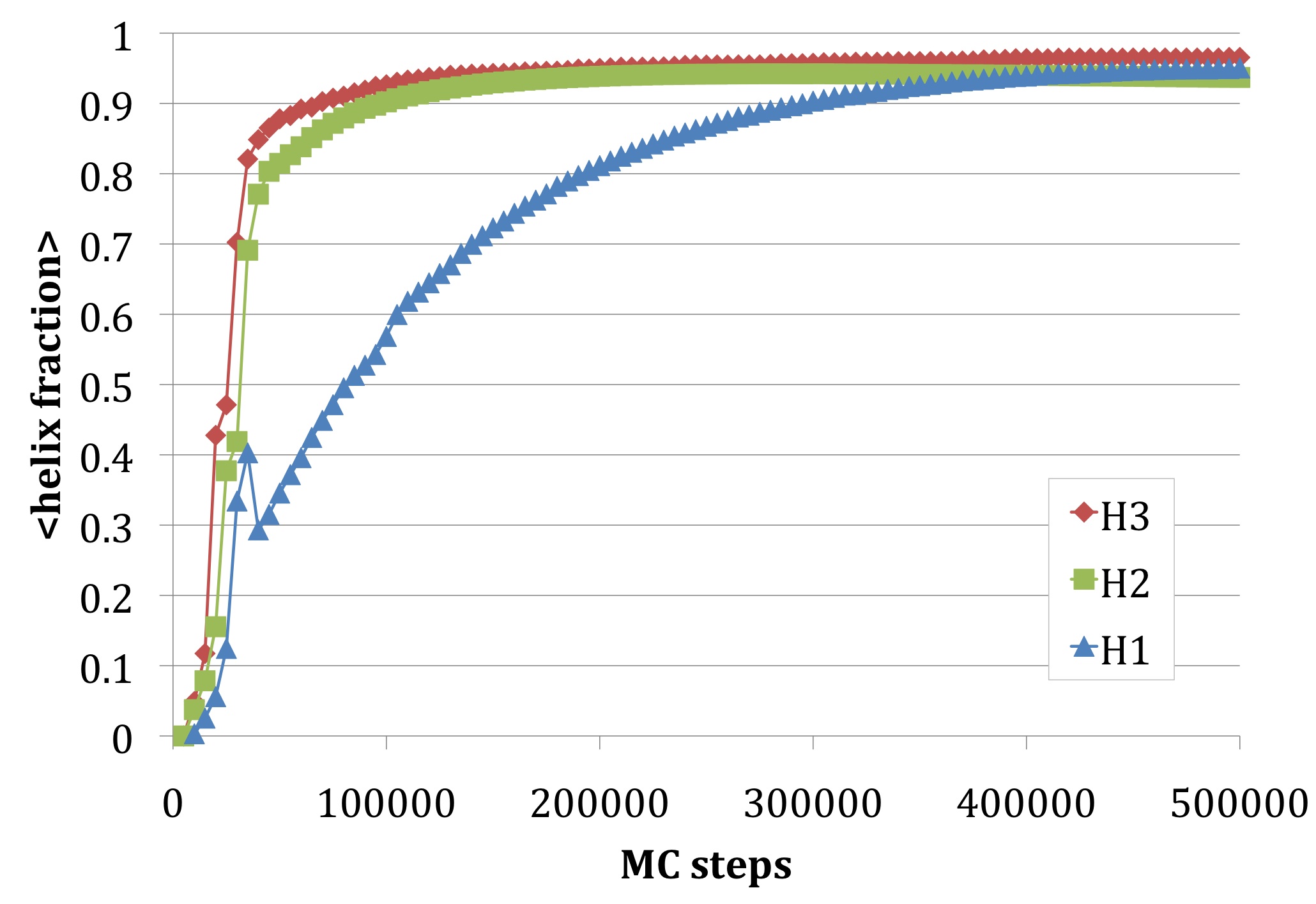}
        \caption{{ \it
        The formation of helices during the collapse of 1YRF, averaged over 4000 trajectories following
        \cite{wallin}. 
        The helix II is formed first, and towards the end of the
        collapse the helix II adjusts itself with the loops. 
        }}
       \label{Figure 1}
\end{figure}
In Figure 2 we display three generic  snap-shot configurations together 
with the final fold. In a typical simulation we arrive at a structure that deviates from
the 1YRF in PDB around 0.5 \.Angstr\"om in RMSD distance. In average, the C$_\alpha$ carbons of
the PDB configuration  has B-factors that correspond to a Debye-Waller fluctuation distance around 0.4 \.A. 
Consequently we can entirely attribute the average RMSD distance of $\sim 0.5$\.A between our final configurations 
and the PDB structure  to  thermal fluctuations.
%
%
%
%
%
\begin{figure}[tbh]
        \centering
                \includegraphics[width=0.45\textwidth]{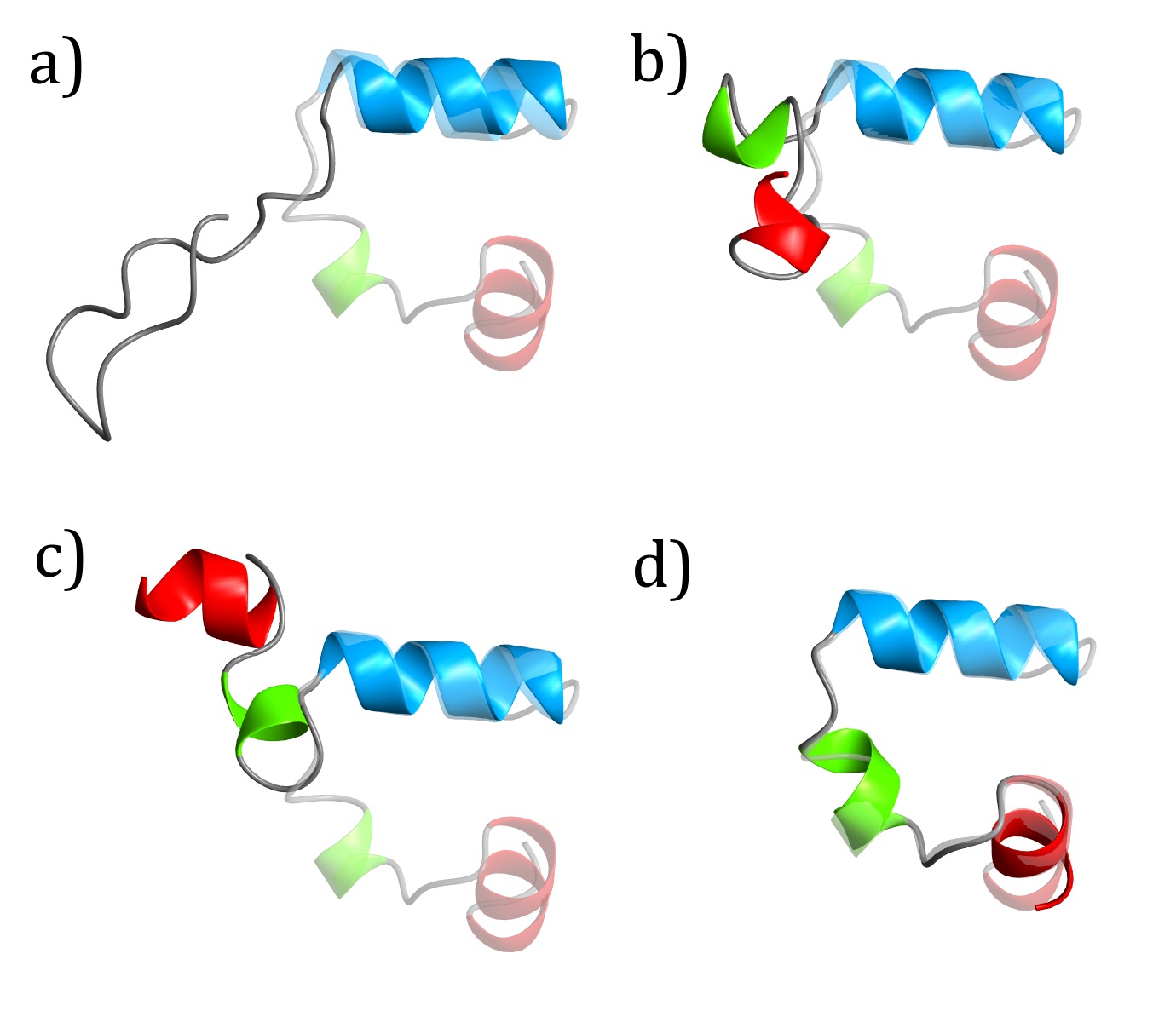}
        \caption{{ \it A set of snap-shots of a generic folding pathway in our simulation, over the shadow of the
        PDB configuration. Color coding shows how the three helices are formed.
     a) The folding starts with formation of helix III. 
     b) After this, there is formation of the other two helices and loops.
     c) The folding proceeds with stabilization of middle helix and loops.
     d) The final fold and the PDB structure of 1YRF generically coincide with RMSD accuracy of about 0.5\.A.
            }}
       \label{Figure 2}
\end{figure}

\section{Discussion} 
The protein folding problem endures as the pre-eminent unresolved conundrum in science.
The major problem in any atomic level  simulation relates to time scales and high energy barriers.
These can only be overcome with coherent multi-atom collective motions, whose molecular dynamics description remains 
a formidable task.  Here we have introduced a new paradigm for describing and modeling 
protein collapse. We propose to address the large scale hurdles in terms of solitons in an effective Landau theory
and to describe the ensuing collapse dynamics universally, by a non-conservative Markovian heat-bath evolution.   
We have demonstrated that in the case of 1YRF where MD simulations are available for comparison, 
the folding pathways obtained in our approach are the same.  Among the future challenges is to
compute the soliton profiles  directly from the individual amino acids, to facilitate predictive collapse simulations.

\end{document}